\documentclass[twocolumn,superscriptaddress,showpacs,preprintnumbers,amsmath,amssymb]{revtex4}

\newcommand{\bb}{\begin{equation}}
\newcommand{\en}{\end{equation}}

\def\vx{{\bf x}}

\usepackage{graphicx}

\begin{document}

\draft

\title{Membrane tension lowering induced by protein activity}
\author{M. D. El Alaoui Faris}
\affiliation{Institut Curie, Centre de recherche; CNRS, UMR 168;
Universit\'e Pierre et Marie Curie, Paris, F-75248 France}

\author{D. Lacoste}
\affiliation{ESPCI, Laboratoire de Physico-Chimie Th\'eorique;
CNRS, UMR 7083, 75231 Paris Cedex 05, France}

\author{J. P\'{e}cr\'{e}aux}
\altaffiliation[Also at ]{MPI for Molecular Cell Biology and
Genetics}
\author{J-F. Joanny}
\affiliation{Institut Curie, Centre de recherche; CNRS, UMR 168;
Universit\'e Pierre et Marie Curie, Paris, F-75248 France}

\author{J. Prost}
\affiliation{Institut Curie, Centre de recherche; CNRS, UMR 168;
Universit\'e Pierre et Marie Curie, Paris, F-75248 France}
\affiliation{ESPCI; ParisTech, 10 rue Vauquelin, 75231 Paris Cedex
05, France}

\author{P. Bassereau}
\affiliation{Institut Curie, Centre de recherche; CNRS, UMR 168;
Universit\'e Pierre et Marie Curie, Paris, F-75248 France}

\begin{abstract}
We present measurements of the fluctuation spectrum of giant
vesicles containing bacteriorhodopsin (BR) pumps using
video-microscopy. When the pumps are activated, we observe a
significant increase of the fluctuations in the low wavevector
region, which we interpret as due to a lowering of the effective
tension of the membrane.
\end{abstract}

\date{\today}

 \maketitle

Membranes are self-assembled bilayers of surfactants or
phospholipids, which form flexible surfaces. The mechanical
properties of membranes are essentially controlled by parameters,
such as the membrane tension, bending modulus and spontaneous
curvature \cite{seifert}. These parameters completely characterize
the membrane fluctuation spectrum at thermal equilibrium, but they
are not sufficient to characterize real biological membranes, such
as cell plasma  membranes, which are non-equilibrium systems
\cite{cell,madan_CRAC}. Biological membranes are in general active
in the sense that they are constantly maintained out of
equilibrium either by active proteins inside the membrane that are
often  ATP-consuming enzymes or by an energy flow, for example a
lipid flux). Due to the complexity of active biological systems
{\em in vivo}, recent studies have focused on {\em in vitro}
biomimetic systems. An example of an active system is a giant
unilamellar vesicle (GUV) which is rendered active by the
inclusion of light-activated bacteriorhodopsin (BR) pumps. The
bacteriorhodopsin pumps transfer protons unidirectionally across
the membrane as a consequence of their conformational changes,
when they are excited by light of a specific wavelength.
Experiments on giant unilamellar vesicles containing
bacteriorhodopsin pumps or Ca$^{2+}$ATPase pumps have shown that
the non-equilibrium forces arising from ion pumps embedded in the
membrane are able to significantly enhance the membrane
fluctuations \cite{RTP,jb_PRE,jb_PRL,girard}. In these
experiments, the membrane tension has been fixed by means of
micropipet aspiration and the corresponding excess area measured.
The slope of such a curve defines an effective temperature,
measuring the enhancement of the membrane fluctuations by the
protein activity. The amplification of the fluctuations due to
activity has been originally predicted in Ref.~\cite{PB}.

In this Letter, we report the first experimental measurements of
an active fluctuation spectrum of GUVs containing
bacteriorhodopsin pumps using video microscopy. The details of the
technique and of the analysis are given in Ref.~\cite{j_pecreaux}.
These measurements are complementary to previous experiments
using micropipets since they probe a range of wavevectors not accessible
otherwise. Micropipet experiments
provide information on the active fluctuations integrated over all
wave vectors which require an independent calibration of the
membrane tension. On the contrary, video microscopy gives a direct
measurement of the fluctuation spectrum from which the membrane
tension can be extracted \cite{j_pecreaux}. In this paper, we
analyze theoretically the active fluctuation spectrum using the
model of Ref.~\cite{lomholt}, which is an extension of the
hydrodynamic model of Ref.~\cite{jb_PRE} to quasi-spherical
vesicles taking into account the active noise of the pumps and
localized active forces.

Let us first discuss the passive behavior of giant vesicles. For
the sake of simplicity we use notations appropriate to
quasi-planar membranes. We consider a quasi-planar membrane
surface in the Monge gauge, in which it is defined by its height
$u(\vx)$ at position $\vx$. Two density fields are defined on the
surface, $\psi^\uparrow$ and $\psi^\downarrow$, corresponding to
the surface densities of proteins oriented with the two possible
orientations \cite{jb_PRE}. The membrane free energy is a function
of the height field $u(\vx)$ and of the imbalance of protein
densities $\psi(\vx)=\psi^\uparrow(\vx)-\psi^\downarrow(\vx)$. We
expand the free energy to quadratic order in these variables:
\begin{equation}\label{free energy}
F[u,\psi]= \frac{1}{2} \int d^2 \vx [ \kappa \left( \nabla^2 u
\right)^2 + \sigma \left( \nabla u \right)^2+ \chi \psi^2 -2 \Xi
\psi \nabla^2 u ],
\end{equation}
where $\kappa$ is the bending modulus, $\sigma$ the surface
tension, $\chi$ a susceptibility coefficient and $\Xi$ a
coefficient characterizing the coupling between the membrane
curvature and the average orientation of the proteins. In the
following, we neglect the curvature induced coupling; this is
justified for bacteriorhodopsin pumps since the incorporation of
 proteins does not lead to any measurable
renormalization of the bending modulus \cite{jb_PRE}. Note that
for dilute proteins in the membrane, $\chi=k_B T/n_0$, where $n_0$
is the average density of proteins, of the order of
$10^{16}$m$^{-2}$. In our experiments, only the vesicle contour at
the equator corresponding to a slice of the vesicle in the plane
$y=0$ is recorded and is used to calculate the fluctuation
spectrum: the measured quantity is
\begin{equation} \label{cut}
\langle |u(q_x, y=0)|^2 \rangle =
\frac{1}{2\pi}\int_{-\infty}^{+\infty} \langle
|u(q_x,q_y)|^2\rangle dq_y.
\end{equation}
The passive equilibrium value of this fluctuation spectrum
is calculated using the equipartition theorem from
Eqs.~\ref{free energy}-\ref{cut}:
\begin{equation}
\langle|u(q_x,y=0)|^2\rangle = \frac{k_BT}{2\sigma}\left[
\frac{1}{q_x} -
\frac{1}{\sqrt{\frac{\sigma}{\kappa}+q_x^2}}\right]=
g(q_x,\sigma,\kappa).
\label{spec-passif-coupe}
\end{equation}
Fluctuation spectra are often analyzed in terms of power laws
defined by $g(q_x,\sigma,\kappa) \sim q_x^\nu$. Two regimes can be
distinguished depending on the position of the wave vector $q_x$
compared to the cross-over wavevector $q_c=\sqrt{\sigma/\kappa}$.
For $q \ll q_c$, membrane tension dominates and $\nu=-1$, whereas
for $q \gg q_c$ bending elasticity dominates and $\nu=-3$. Our
experiments lead to $\nu \simeq -2$, in the cross-over region
between these two regimes.

We now discuss active membrane fluctuations. In
Ref.~\cite{jb_PRE}, a hydrodynamic theory was developed to
calculate the non-equilibrium fluctuations of an active membrane
containing ion pumps. This work has stimulated substantial
theoretical interest, focused mainly on the general question of
the proper description of non-equilibrium effects associated to
protein conformation changes \cite{gautam,gov}. More recent
developments of a similar hydrodynamic approach have lead to a
general theoretical description of active gels
\cite{simha,madan_active_filament,joanny}. In
Ref.~\cite{lomholt}, the active force distribution is modeled as a
superposition of dipoles located along the membrane normal ${\bf
n}$ (and not along $z$ as in previous models), which generalizes
Ref.~\cite{jb_PRE}. Experiments do not probe this force
distribution directly but rather its first two moments. The first
moment represents the active contribution to the membrane tension and is
denoted by $\sigma_{dip}$ and the second moment $Q$ represents the
modification of the membrane bending moments
by the activity of the force dipoles \cite{lomholt}. Both active contributions to the
tension and to the membrane curvature are also present in
Ref.~\cite{salbreux} where the activity is due the the myosin
molecular motors in the cortical layer bound to the cell membrane.

The fluctuation spectrum of an active membrane in a
quasi-spherical geometry can be calculated using an expansion
around a spherical shape with radius $R_0$ to first order in the
deviations from the sphere using spherical harmonics, with ${\bf
R}={\bf R_0} + {\bf n} R_0 \sum_{l,m} u_{lm} \mathcal{Y}_{lm}$.
The fluctuation spectrum of a quasi-spherical vesicle is
characterized by the amplitude of the spherical modes $\langle
|u_{lm}|^2 \rangle$ which are function of $l$ only for symmetry
reasons. In Ref.~\cite{j_pecreaux}, the fluctuation spectrum of a
quasi-spherical vesicle is compared to the corresponding spectrum
for a planar membrane, and this comparison showed that for modes
$l>5$, the fluctuation spectrum of a quasi-spherical vesicle are
very well approximated by that of a planar membrane at a wave
vector $q$, $\langle |u(q)|^2 \rangle$; at a wavevector $q=l/R_0$,
the two spectra are related by $\langle |u(q)|^2 \rangle=R_0^4
\langle |u_{lm}|^2 \rangle$. For the sake of simplicity, we
discuss in the following the effect of the activity on the
fluctuation spectrum only at the level of planar membranes. We
have also simplified the calculation of Ref.~\cite{lomholt}, by
neglecting the shot-noise, which is the intrinsic noise of the
pumps \cite{PB}. This approximation is reasonable because the main
effect of the shot-noise is to modify the fluctuation spectrum in
the high wavevector region \cite{lomholt} which is not the region
where we observe a large modification of the spectrum due to
activity in our experiments. We denote here $F_2$ the constant entering in
$Q=F_2 \psi$  \cite{lomholt}. This generalizes the notation of
Ref.~\cite{jb_PRE}, where this quantity represented the
quadrupole moment of the force dipole. With these three approximations, we
obtain the following spectrum
\begin{equation}
\langle |u(q_x, q_y)|^2 \rangle=\frac{k_BT}{\kappa q^4 +
\tilde{\sigma} q^2}\biggl\{ 1 +
\frac{F_{2}^2}{4\chi}\times\frac{q^2}{\kappa q^2 + 4\eta D q +
\tilde{\sigma} } \biggr\}, \label{SpecLomholtp1}
\end{equation}
where $D$ is the diffusion coefficient of
the proteins which is of
the order of $10^{-12}$m$^2$s$^{-1}$ and $\eta$ is the viscosity
of water. Since the tension of the vesicles is at least of the
order of $10^{-8}$N/m, we find that the term $4 \eta D q$ can be
neglected with respect to $\tilde{\sigma}$ in the range of
wavevectors relevant to the experiments \cite{kahya}.
The effective tension $\tilde{\sigma}$ is defined as
$\tilde{\sigma}=\sigma + \sigma_{dip}$, where $\sigma$ is the
passive contribution to the tension and $\sigma_{dip}$
the active
part due to BR proteins. After integration as in
Eq.~\ref{spec-passif-coupe}, we find
\begin{equation}
\langle |u(q_x, y=0)|^2 \rangle =g(q_x,\tilde{\sigma},\kappa) +
 \frac{F_{2}^2 n_0}{16\kappa^2}\times \frac{1}{(q_x^2 +
 \frac{\tilde{\sigma}}{\kappa})^{3/2}},
\label{SpecLomQ}
\end{equation}
in which both terms on the right hand side contain active
contributions when $\sigma_{dip}$ or $F_2$ do not vanish.
Note that
Eq.~\ref{SpecLomholtp1} is very similar to the active spectrum
given Ref.~\cite{jb_PRE} with the correspondence $F_2=2 w
\mathcal{P}_a$ except for one important difference: the
tension contains an active contribution but not the bending
modulus whereas in Ref.~\cite{jb_PRE} it is the opposite.
Because of this difference the active spectrum of
Ref.~\cite{jb_PRE} does
not fit our experiments (it would predict a large effect
of activity
at high wavevectors whereas the effect is observed here mainly at small
wavevectors). Another consequence of this difference, is
that the active spectrum of Eq.~\ref{SpecLomQ} cannot in
general be described in terms of an active temperature even
in the low wavevector limit \cite{weikl}.

\begin{figure}
{\par {
\rotatebox{0}{\includegraphics[scale=1]{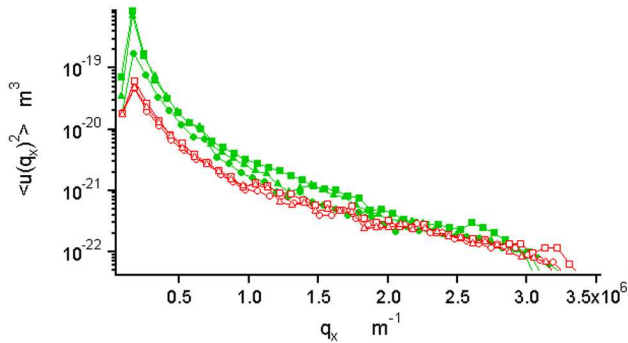}} }
\par}
\caption{Fluctuation spectrum of a single GUV containing BR in a
buffer with 1mM sodium azide. The red (empty) symbols correspond
to three consecutive recorded passive spectra, and the green
(filled) ones to three consecutive active spectra.}
\label{Fig:spectrum}
\end{figure}

Experiments have been performed with GUVs made of
Egg Phosphatidyl
Choline at a molar fraction of 240 lipids per protein, using
the BR reconstitution protocol developed in Ref.~\cite{girard2}. We have
used a PEG passivated substrate and checked that vesicles fluctuate without
adhering \cite{j_pecreaux}. In Fig.~\ref{Fig:spectrum}, we show the measured
fluctuation spectrum
for the same GUV containing BR in active and passive states.
To ensure reproducibility of the data, three consecutive passive
and then, three active spectra were recorded. This figure confirms that
activity leads to an enhancement of the fluctuations as expected
from previous experimental studies
\cite{jb_PRE,jb_PRL}. It is important to point out that
to observe
this effect 1mM of azide must be present in the solution. In the
absence of this compound no enhancement of the
fluctuation spectrum was observed \cite{giahi}.
This observation is consistent with several studies
which suggest
that azide enhances the proton transfer in BR \cite{oesterhelt}.
A large enhancement of the active as compared
to passive fluctuations is observed at low wavevectors, which we
attribute to a lowering of the membrane tension due to
the activity.

\begin{figure}
{\par {
\rotatebox{0}{\includegraphics[scale=1]{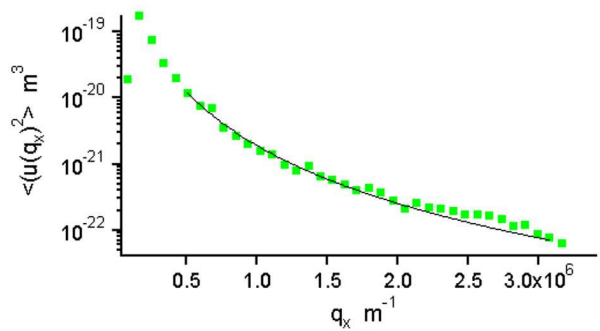}} }
\par}
\caption{Fit of one active spectra of Fig.~\ref{Fig:spectrum}
using Eq.~\ref{SpecLomQ}. The parameter apparent $\kappa$ was
fixed to $5.6 \cdot 10^{-19}$J obtained by fitting the passive
spectrum. This fit gives
$\tilde{\sigma}=5.3 \cdot 10^{-8}$N/m and
$F_2=3.9 \cdot 10^{-28}$ Jm. } \label{Fig:fit-spectrum}
\end{figure}

We first fit the passive
spectrum using Eq.~\ref{spec-passif-coupe}. This leads to
$\sigma=3.9 \cdot 10^{-7} \pm 3 \cdot 10^{-8}$N/m and an apparent
bending modulus $\kappa=5.6 \cdot 10^{-19} \pm 4.4 \cdot
10^{-20}$J. The value of the bending modulus deduced from such a fit is
overestimated. Several artifacts perturb the measurements at high
wavevectors: the pixel noise due to the discrete detection of the
images, the integration time of the camera, and the effect of
gravity \cite{j_pecreaux}.
Taking into account these corrections at high $q$ becomes very
difficult in the active case, we thus have chosen a simple fitting
procedure, in which the corrections due to the integration time
and to the gravity are neglected for both passive and active spectra.
This procedure is justified by the observation that both
spectra are superimposed at high $q$.
The determination of the bending modulus
$\kappa$  is therefore delicate with
our video microscopy technique while the measurement of
the tension is reliable.
With the value of the bending
modulus kept fixed, we fit the three active spectra of
Fig.~\ref{Fig:spectrum} using Eq.~\ref{SpecLomQ}. One fit which is
shown in Fig.~\ref{Fig:fit-spectrum} leads to $\tilde{\sigma}=5.3
\cdot 10^{-8}$N/m and $F_2=4 \cdot 10^{-28}$Jm. Thus the active
contribution to the tension $\sigma_{dip}=\tilde{\sigma}-\sigma$
is negative and of the order of $-3.3 \cdot 10^{-7}$N/m for this
particular vesicle (see table I for a summary of the results of
the fits for the GUV of Fig.~\ref{Fig:spectrum}). The lowering of
the tension is systematic in all our experiments. The fact that
$\sigma_{dip} \simeq -3.3 \cdot 10^{-7}$N/m is very different from
the estimate of $1.6 \cdot 10^{-3}$N/m based on
Refs.~\cite{lomholt,jb_PRE} suggests that the modeling of the
force distribution of these references is not appropriate to our
experiments. 
In table I, we have also shown the exponent $\nu$,
which is the apparent power law exponent of the spectrum. When the
vesicle is active the exponent gets closer to $-3$, which is the
expected value in the bending dominated regime. This is consistent
with a lowering of the cross-over wavevector
$q_c=\sqrt{\tilde{\sigma}/\kappa}$, and thus with a lowering of
the tension since in this model $\kappa$ is not affected by
activity. 
We have ignored any dependence of the protein activity on the local membrane curvature, an assumption justified by the observation that the main effect is a correction to the tension and not to the bending modulus.
The value of the parameter $F_2$ deduced from our fit is
of the same order of magnitude as that estimated in
Ref.~\cite{jb_PRE} based on micropipet experiments since the
factor 2-3 increase in effective temperature measured in this
reference leads to $F_2 \simeq 9 \cdot 10^{-28}$Jm. 
To summarize, Eq.~\ref{SpecLomholtp1} successfully describes both the micropipet and the
fluctuation spectrum measurements. 

We now discuss in more details the lowering of the
membrane tension due to activity. We have checked that
this effect is
compatible with the constraints of constant surface and
volume of
the vesicle, which must be imposed if permeation is
negligible \cite{jb_PRE}. To do so, we estimate the excess
area $\alpha=\int q^3 dq \langle |u(q)|^2 \rangle /4/\pi$ for
passive and active vesicles based on the above fluctuation
spectra, by taking the lower bound of integration to
be $q_{min}=2 \pi/R_0$, with $R_0=10 \mu$m the radius of
the vesicles; this lower bound corresponds roughly
to the lowest measured value
in Fig.~\ref{Fig:spectrum}. In the accessible range of
$q$ values,
the excess area density $q^3 \langle |u(q)|^2 \rangle /4/\pi$
of the active vesicles is always larger than the excess area
density of the passive vesicles. At higher $q$ wavevectors,
it should be the
opposite; but the $q$ range where this would happen may be
outside the accessible range.

\begin{table}[h!]
\begin{center}
\begin{tabular}{|l|c|c|c|}
\hline
Spectrum & $\sigma$ (x $10^{-7}$ N/m) & $F_{2}$ (x $10^{-28}$ J.m) & $\nu$ \\
\hline
passive & $3.9 \pm 0.3$ & 0 (imposed) & $-2.15 \pm 0.04$ \\
\hline
active 1 & $0.53 \pm 0.2$ & $3.9 \pm 1.1$ & $-2.7 \pm 0.04$ \\
active 2 & $0.79 \pm 0.06$ & $9.6 \pm 1.1$ & $-2.78 \pm 0.06$ \\
active 3 & $0.35 \pm 0.1$ & $7.2 \pm 0.6$ &  $-2.87 \pm 0.07$ \\
\hline
\end{tabular}
\caption{Fitting parameters using the fluctuation spectrum of
Eq.~\ref{SpecLomQ}. The passive spectrum is fitted with the
condition $F_{2}=0$. Then three consecutive active spectra are
fitted with the same vesicle using the value of $\kappa_{ap}=5.6
\cdot 10^{-19} \pm 4.4 \cdot 10^{-20}$J determined from the fit of
the passive spectrum. From this, the membrane tension $\sigma$,
$F_{2}$ and the exponent $\nu$ are deduced.}
\label{tabFa}
\end{center}
\end{table}

Another important question is the sign of the effect, which can
not be fixed by symmetry arguments \cite{RTP}. Indeed, both
orientations of the force dipoles \cite{gautam,jb_PRE}, inwards or
outwards with respect to the membrane surface, are possible. The
present experiments suggest that here the force dipoles must have
an inward orientation (corresponding to contractile forces
\cite{madan_active_filament}) to produce a lowering of the
tension. Another possibility would be to consider lateral dipoles.
Here, we propose that the lowering of the tension is a consequence
of electrostatic effects. This interpretation is not incompatible
with the fact that BR conformational changes also play an
important role. We have checked experimentally that a pH gradient
builds up in our GUVs when they are activated, which suggests that
a significant voltage drop exists across the membrane. Furthermore
in membranes containing BR, the activity of the BR can be
suppressed by applying an external voltage drop across the
membrane \cite{geibel}. In view of these observations, we propose
that Maxwell stresses created by ion transport lead to a
renormalization of the tension as shown in a simple model
developed by two of us \cite{lacoste}. In this model, the
electrostatic correction to the tension is proportional to the
square of the electric current going across the membrane. Maxwell
stresses are large because of the
large mismatch in dielectric constants at the membrane, and they tend to
reduce the membrane area, thus producing a negative electrostatic
contribution to the tension. A lowering of the membrane tension
due to the application of normal electric fields very generally
leads to instabilities of the membrane
\cite{dimova,lomholt_elect}.

To summarize, we have presented what are to our knowledge the
first measurements of active fluctuation spectra of GUVs
containing BR proteins using video microscopy. The experiments
show that activity enhances the fluctuations. The effect is strong
in the low wavevector region, which we interpret as a lowering of
the tension due to activity.
This effect could not have been detected in micropipet experiments which are
done at constant membrane tension.
A possible candidate for explaining
the lowering of the tension would be that it is caused by Maxwell
stresses due to the transport of ions across the membrane. Further
experiments (using for instance patch-clamp techniques with the
same GUV or other GUVs containing ion channels
rather pumps) are required to confirm these suggestions. \\

\begin{acknowledgments}
We acknowledge many stimulating discussions with J. Tittor, G.
Menon, S. Ramaswamy, N. Gov and M. Lomholt. We thank D. Oesterhelt
for kindly providing us with BR. We also acknowledge support from
the European Network of Excellence SoftComp (NMP3-CT-2004-502235)
and
 the Indo-French Center CEFIPRA (grant 3504-2).
 \end{acknowledgments}

\bibliographystyle{apsrev}
\bibliography{membranes}
\end{document}